\documentclass[amsmath,amssymb,prb,floatfix,preprint,superscriptaddress]{revtex4-2}

\usepackage{graphicx}
\usepackage{dcolumn}
\usepackage{bm}
\usepackage{hyperref}
\usepackage{siunitx}
\usepackage{soul} 
\usepackage{ulem}

\graphicspath{{}, {/figures/}}

\begin{document} 

\title{Hybrid cavity-antenna architecture for strong and tunable sideband-selective molecular Raman scattering enhancement} 

\author{Ilan Shlesinger}
\affiliation{Department of Information in Matter and Center for Nanophotonics, AMOLF, Science Park 104, 1098 XG, Amsterdam, Netherlands}
\affiliation{Mat\'{e}riaux et Ph\'{e}nom\`{e}nes Quantique, Universit\'{e} Paris Cit\'{e}, CNRS UMR 7162, Paris, France}
\author{Jente Vandersmissen}
\affiliation{Department of Information in Matter and Center for Nanophotonics, AMOLF, Science Park 104, 1098 XG, Amsterdam, Netherlands}
\author{Eitan Oksenberg}
\affiliation{Department of Information in Matter and Center for Nanophotonics, AMOLF, Science Park 104, 1098 XG, Amsterdam, Netherlands}
\author{Ewold Verhagen}
\affiliation{Department of Information in Matter and Center for Nanophotonics, AMOLF, Science Park 104, 1098 XG, Amsterdam, Netherlands}
\author{A. Femius Koenderink}
\affiliation{Department of Information in Matter and Center for Nanophotonics, AMOLF, Science Park 104, 1098 XG, Amsterdam, Netherlands}
 
\date{\today}

\begin{abstract}
Plasmon resonances at the surface of plasmonic antennas allow for extremely strong enhancement of Raman scattering.  Intrinsic to plasmonics, however, is that extreme field confinement lacks precise spectral control, which would hold great promise in shaping the optomechanical interaction between light and molecular vibrations at will.
We demonstrate an experimental platform composed of a plasmonic nanocube-on-mirror antenna coupled to an open, tunable Fabry-Perot microcavity for selective addressing of individual vibrational lines of molecules with strong Raman scattering enhancement. Multiple narrow and intense optical resonances arising from the hybridization of the cavity modes and the plasmonic broad resonance are used to simultaneously enhance the laser pump and the local density of optical states (LDOS) and are characterized using rigorous modal analysis. The versatile bottom-up fabrication approach permits quantitative comparison with the bare nanocube-on-mirror system, both theoretically and experimentally. This shows that the hybrid system allows for similar SERS enhancement ratios with narrow optical modes, paving the way for dynamical back action effects in molecular optomechanics.
\end{abstract}
\maketitle

\section{Introduction}
Surface-enhanced Raman spectroscopy (SERS)~\cite{Jeanmaire1977,Otto1992,pilot_review_2019}  is a powerful spectroscopy method that relies on the extreme field confinement of metallic plasmonic surfaces to strongly enhance Raman scattering by a molecule~\cite{Moskovits2005,Willets2007}.
Recently, SERS has been described within the framework of cavity optomechanics \cite{Roelli2016,Esteban2022,Schmidt2017LinkingSERS}. 
This viewpoint gives a picture of the optical forces of the plasmon resonator mode onto the molecule's mechanical modes, providing an explanation for various proposed and observed effects such as parametric cooling or amplification of the mechanical motion, and hence nonlinear SERS enhancements~\cite{Benz2016Single-moleculepicocavities,Lombardi2018,Xu2022}, as well as coherent infrared-visible conversion~\cite{Chen2021,Xomalis2021}. Most of the promises contained in this vista of cavity optomechanics transposed to molecular vibrations require the sideband-resolved regime, where the optical resonance linewidth is narrow compared to the mechanical vibrational frequency of the molecules. This allows controlling the phase of the optomechanical interaction, selectively enhancing or suppressing Stokes or anti-Stokes scattering processes through dynamical backaction of the optical resonance~\cite{Aspelmeyer2014CavityOptomechanics}. Because of the large decay rate of plasmonic resonances, the electromagnetic enhancement offered in SERS is intrinsically broad, limiting the coherence of the optomechanical interaction. While the low quality factors (Q) of plasmonics mean that sideband-resolved plasmonic SERS is out of reach, recent work has argued that sideband-resolved SERS is possible via the use of a hybrid plasmonic-photonic resonator, both theoretically \cite{Dezgouli2019,Shlesinger2021} and experimentally \cite{Shlesinger2023}. A hybrid resonator combines a high-Q photonic resonator with a low mode volume plasmonic resonance~\cite{Shopova2008,Deangelis2008,Doeleman2016,Gurlek2018ManipulationStrong-Coupling}. The resulting resonances are narrower than the usual molecular vibrational resonance frequencies allowing for sideband-resolved SERS with single optical resonators and strong SERS enhancements~\cite{Dezfouli2017,Shlesinger2021}.\\ 
The single realization to date, however, \cite{Shlesinger2023} relied on lithography to define both the resonator and the plasmonic antenna. This not only means that there is no tunability after fabrication, but also that the properties of the very best plasmonic SERS structures cannot be exploited, such as ultra-narrow gaps and atomically controlled metallic facets~\cite{Baumberg2019}.

In this paper, we report selective Raman enhancement in a new generation of hybrid resonators that combine state of the art gap antenna modes~\cite{Xomalis2020} together with a tunable Fabry-P\'erot (FP) cavity~\cite{Barbour2011}. This system allows reaching record SERS enhancements in a sideband resolved regime, with the possibility to tune the hybrid mode resonance frequency in situ, and thus independently and selectively enhance single Raman lines.  The gap antenna is made directly on one of the two FP mirrors, and is essentially the well known nanocube-on-mirror geometry with all its electromagnetic and structural advantages~\cite{Moreau2012,Xomalis2020}. The bottom-up fabrication approach that we adopt allows for a full comparison of the hybrid cavity performance with the bare constituents from which it is made, and in particular with the bare nanocube-on-mirror system. We show that the tunable hybrid SERS performance is at least equivalent in strength to the one of the ideal nanoantenna system, but accessible with much lower NA optical components, due to efficient funneling into the FP cavity mode.

\section{Results}
\begin{figure}
    \centering
    \includegraphics[width=0.6\textwidth]{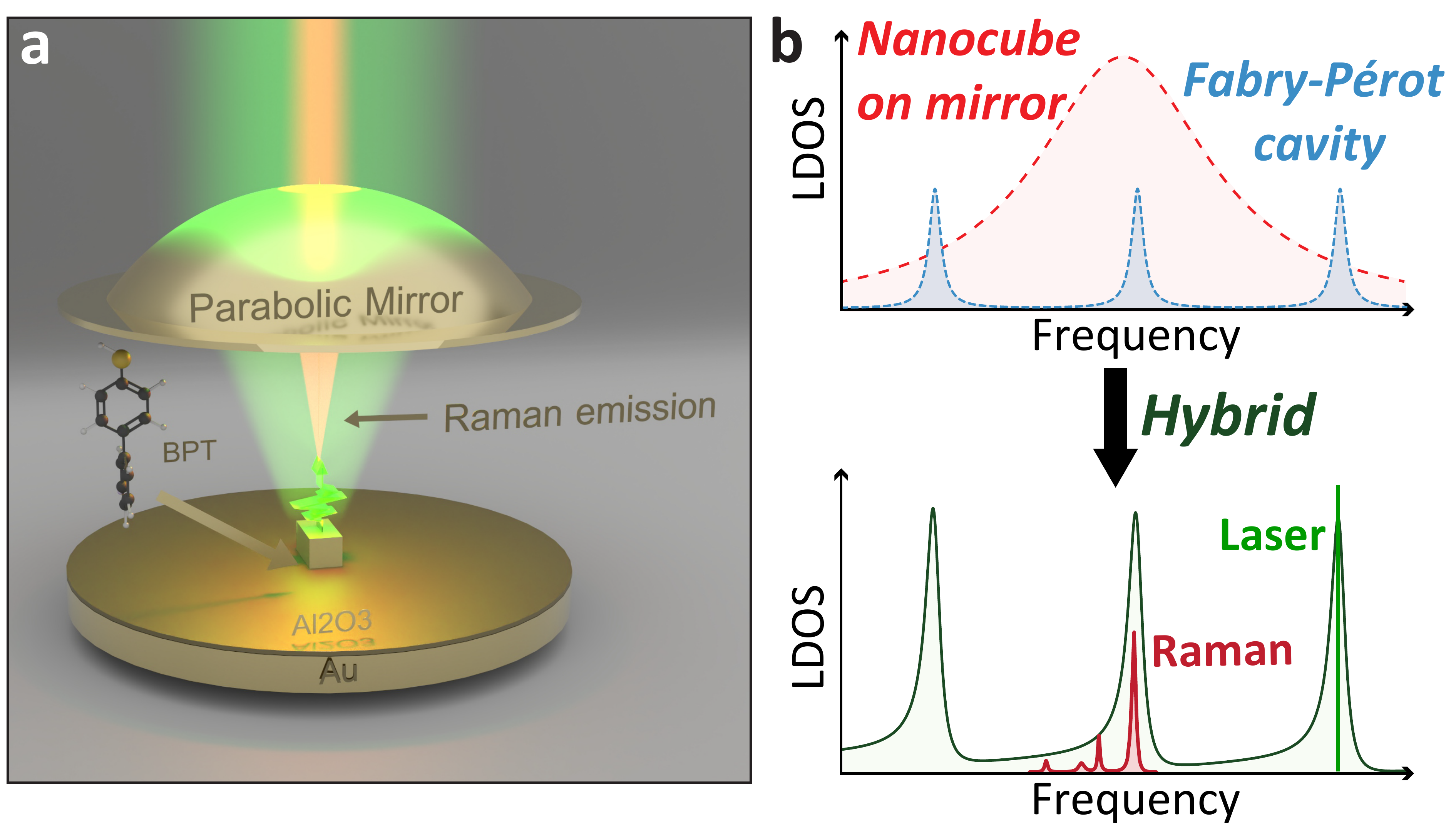}
    \caption{\textbf{a} SERS with a hybrid resonator that combines a nanocube on gold embedded in a tunable FP cavity. The nanocube is coated with byphenol-4-thiol molecules, whose Raman emission can be tuned by changing the cavity opening length. \textbf{b} Schematic representation of the SERS enhancement obtained with the hybrid resonator, offering high LDOS and sideband resolution.
}
    \label{fig:Concept}
\end{figure}

We realize the system as depicted in Figure \ref{fig:Concept} as a tunable hybrid cavity, displaying both a strong field confinement and a high quality factor. The system consists of a monocrystalline gold nanocube on top of a flat gold mirror forming a nanocube-on-mirror (NCoM) system~\cite{Moreau2012}, embedded in a tunable, open-access Fabry-P\'erot microcavity~\cite{Kelkar2015}. 
The NCoM system exhibits an in-plane magnetic dipole mode that radiates vertically away from the mirror \cite{Lassiter2013,Chikkaraddy2017}. This mode has a broad resonance and exhibits very high local density of states (LDOS) in the middle of the gap.
The gold cube is separated from the mirror by a thin Al$_2$O$_3$ dielectric spacer layer whose thickness can be tuned (1 to 15 nm) to control the gap resonance frequency. 
By placing the NCoM in a Fabry-P\'erot cavity, created by a second, concave, mirror, the cavity mode and plasmon antenna mode hybridize, combining the electromagnetic field spatial confinement and LDOS of the NCoM system and the narrow resonance of the Fabry-P\'erot cavity. This hybrid mode realizes mode volumes V$_m$ of around  $10^{-4}\;\lambda^3$ and Q factors of $\approx 300$. The hybrid resonator is used to selectively address individual vibrational modes of molecules. More specifically, monolayers of biphenyl-4-thiol (BPT) are self-assembled on the cube, automatically placing these molecules in the region where the strong field enhancement from the NCoM is present. By virtue of the small linewidth of the hybrid optical mode, it becomes possible to achieve sideband-resolved SERS and have well controlled optomechanical interaction. The optical resonances can be tuned with the cavity opening, so that it can enhance both the pump laser and a single vibrational line. To this end, the free spectral range (FSR) is matched to the vibrational energy of the molecule, such that one cavity mode matches the pump wavelength, and another matches the vibrational line of interest. 

\begin{figure*}
    \centering
    \includegraphics[width=\textwidth]{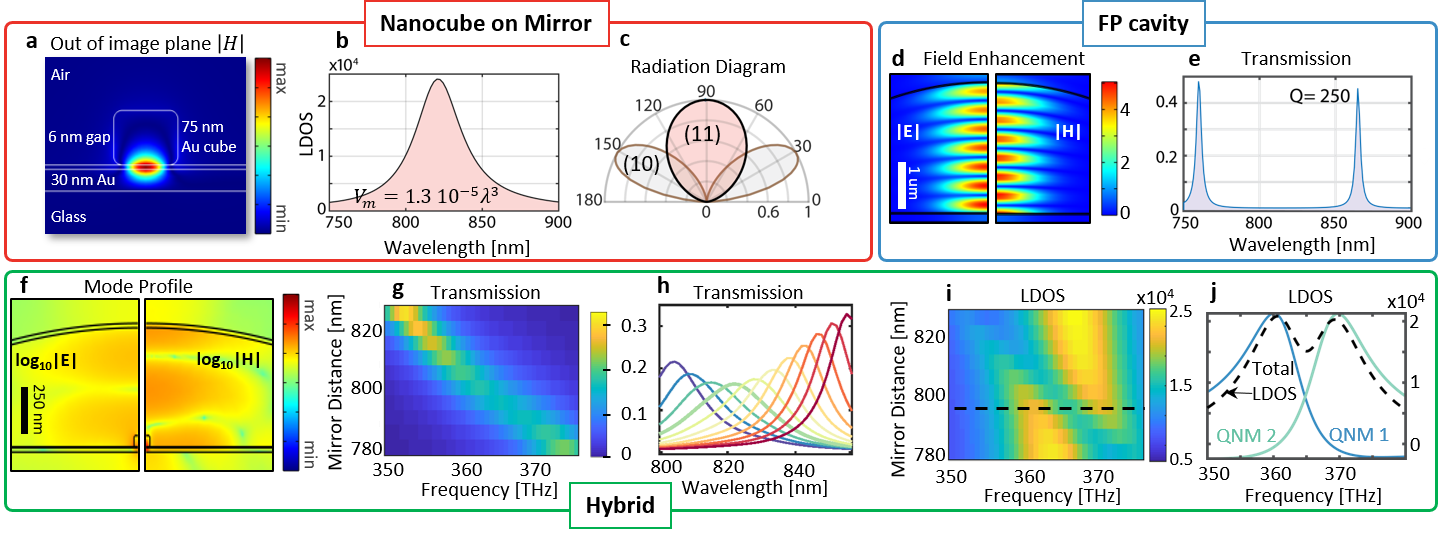}
    \caption{Simulation results obtained with COMSOL and the quasi-normal modal analysis performed with the MAN package~\cite{wu_man_2022}. (\textbf{a-c}) Magnetic mode (11) of the NCoM tuned into resonances by a 6 nm alumina gap, featuring (\textbf{b}) strong LDOS enhancement and (\textbf{c}, red) vertical radiation pattern as opposed to the usual dipolar (10) mode radiation pattern (\textbf{c}, brown). (\textbf{d, e}) A FP cavity is obtained with a second curved mirror, with Qs greater than 200 for 30 nm thick gold mirrors. Both individual components are coupled through their strong H fields on the mirror surface (\textbf{a} and \textbf{d}). The hybridized structure (\textbf{f}) has a peak of tranmission at the (hybrid) cavity resonance (\textbf{g,h}), whose linewidth and resonance frequency changes as a function of the detuning with the NCoM. The total LDOS (\textbf{i} and cross-cut in \textbf{j}) is a sum of the contribution of a mode from each individual resonator, and reaches values up to that of the bare NCoM but with an additional control on the quality factor and resonance frequency.}
    \label{fig:Sim}
\end{figure*}

To elucidate the design characteristics, we discuss COMSOL simulations of first the nanocube-on-mirror system, then the microcavity, and finally the hybrid system. We performed modal analysis to understand the hybridization of the nanocube-on-mirror system with the cavity mode. To this end we employ the so-called quasi-normal mode formalism in the implementation reported by  Wu et al. \cite{wu_man_2022}.\\
The nanocube-on-mirror system supports several modes that have strong field confinement in the gap \cite{Lassiter2013}. Our mode of interest is a (11) mode with an in-plane magnetic dipole mode in the center of the gap, visible in Figure \ref{fig:Sim}a. 
The resonance frequency of this mode can be tuned by varying the spacer layer thickness or the cube dimensions. For an Al$_2$O$_3$ spacer thickness of 6 nm, a cube side length of 75 nm and a semi-transparent 30 nm gold mirror, the LDOS and radiation pattern of the (11) mode are shown in figure \ref{fig:Sim}b and \ref{fig:Sim}c and display a resonance wavelength near 820 nm. The LDOS is calculated near the edge of the cube, 1 nm below the cube, where the molecules are expected to be located. 
LDOS values of up to 24000 are reached at Q = 22, which are expressed in mode volume terms correponds to order $1.3\cdot10^{-5}\;\lambda^3$. The LDOS  has a relatively high radiative contribution of around 34\% despite the nanometer gaps and despite the unconventional choice of just a thin 30 nm Au mirror~\cite{Yang2016}. This thin mirror is chosen such that in experiments we have transmission through the mirror, which enables imaging the location of the nanocubes inside the cavity by looking from outside the cavity through the mirror~\cite{Chikkaraddy2021}. 
The radiation pattern of the (11) mode has a strong lobe perpendicular to the mirror, as shown in Figure~\ref{fig:Sim}(b), with an angular FWHM of 45 degrees. This radiation pattern is well matched to the modest-NA parabolic mirror that we will use to close the microcavity, and is hence advantageous for efficient hybridization with cavity modes.  For reference, the radiation pattern of the vertical dipole mode (10) is also shown. It has a radiation pattern whose maximum is at 60 degrees away from the surface normal, quite unsuited for hybridization with a Fabry-P\'erot mode.

Next, we report on the microcavity properties that one can obtain with a 30 nm thick flat gold mirror and a 30 nm thick parabolic mirror, also from gold and with a radius of curvature of 8 \textmu m. Electric and magnetic field intensities for a typical cavity mode are shown, as well as cavity transmission, in Figs \ref{fig:Sim}d and \ref{fig:Sim}e. We focus on the fundamental transverse modes at low longitudinal mode numbers $m$. The cavity resonances in the near-infrared for a 3 \textmu m cavity opening (yielding a FSR of 110 nm)  feature optical quality factors of  ca. 250 mainly limited by the mirror thickness. The magnetic field experiences a maximum at the mirror surface, naturally matching the in-plane magnetic dipole of the (11) mode of the cube, making it ideal for hybridization.  

The electric and magnetic field profiles of the hybrid resonator mode  combining the NCoM in a FP cavity are shown in Figure~\ref{fig:Sim}f, where we recognize the $m=2$ cavity mode, obtained for a 800 nm opening of the cavity. The hybrid character of the selected mode is demonstrated by the strong field components in the nanogap that are typical of NCoM systems. The LDOS, mode volume, and Q of the hybrid mode depend on the spectral detuning between the NCoM and the cavity resonances. Taking hybrid mode volumes from the quasi-normal mode formalism \cite{Sauvan2013}, we find values on the order of $V_m\sim10^{-4}$ and $Q\sim100$. The hybrid resonator thus will allow reaching sufficiently narrow linewidths to enhance single vibrational modes of the molecule, with only a modest loss in field confinement compared to the NCoM without the closing mirror (one order of magnitude higher mode volume, very similar LDOS values).

The properties of the hybrid resonances are explored by mapping the LDOS at 1 nm below the cube corner as a function of frequency and for different cavity openings (Figure~\ref{fig:Sim}i). The LDOS reaches high values above 10000, of the same order as the values calculated for the bare NCoM and an order of magnitude higher than the state of the art demonstrated hybrid plasmonic-photonic resonators~\cite{Shlesinger2023}.
For high-Q microcavities coupled weakly to dipolar nano-antennas it has been reported that the LDOS follows Fano lineshapes \cite{Doeleman2016,Thakkar2017SculptingHybridization,Shlesinger2021}, evident as very sharp and asymmetric features on the broad backgorund of the plasmon resonances. The diagram obtained for the LDOS in this system may visually suggest an anti-crossing between two resonances, but should not be interpreted as strong coupling. Instead, Fano features appear in the same spirit as in previous work~\cite{Doeleman2020,Shlesinger2023}, with the main difference that the contrast between quality factors is smaller in our system since the cavity Q is only one order of magnitude larger than that of the antenna. This results in comparatively smooth LDOS variations near the cavity frequency.  Figure~\ref{fig:Sim}j shows the LDOS spectrum when the cavity is on resonance with the antenna mode. The system exhibits two maxima in LDOS, which using QNM theory we can decompose as being due to two hybridized eigenmodes.  From a fit of the LDOS 2D map and using a semi-analytical model~\cite{Shlesinger2021} we conclude that there is a hybrid coupling constant $J/2\pi=3.5\,\mathrm{THz}$ to be compared with damping rates $\kappa/2\pi=5\,\mathrm{THz}$ and $\gamma/2\pi=17\,\mathrm{THz}$ for the cavity-like and antenna-like modes. These values indicate that the system is not considered to be in the strong coupling regime, since the optical coupling strength between the two subsystems is still smaller than both of the linewidths. 
The interaction between the two modes results in non-Lorentzian lineshapes for the LDOS, typical of interference effects in dissipative, multimode systems~\cite{Lalanne2018}, which can also display negative contributions of a single mode to the total LDOS. This can be quantified by the complex nature of the mode volume derived from QNM analysis~\cite{Sauvan2013}, which here reaches significant non-real values up to $\mathrm{Im}(V_m)/\mathrm{Re}(V_m)\simeq1$. Interestingly, in this system they arise mainly due to the interaction of the two modes, since for each bare constituent separately (the FP cavity or even the NCoM antenna), the mode volume is mostly real ($\mathrm{Im}(V_m)/\mathrm{Re}(V_m)<0.1$). This demonstrates the possibility of using hybrid resonators to investigate dissipative coupling in molecular optomechanics, where the optical cavity resonance modulates the damping rate of a molecular vibration, which can significantly facilitate reaching high cooperativity regimes~\cite{Primo2020}.

The simulated transmission of an incident Gaussian beam  through the the hybrid system is shown in Figure~\ref{fig:Sim}g,h. When  tuning the cavity frequency over the NCoM resonance frequency, the main effect is a modest  decrease in quality factor. While this observation is in remarkable contrast to  the strong feature in the LDOS map, the  predicted small variation in transmission is  commensurate with cavity perturbation theory \cite{Koenderink2005, Ruesink2015,Kelkar2015}. Cavity perturbation theory predicts a complex-valued cavity frequency shift that is proportional to the antenna polarizability, and which is hence  expected to be fully imaginary for on-resonance coupling. The resulting reduction in Q is mainly visible as a variation in the maximum cavity transmission --- the choice of 30 nm thick mirrors allows reaching good maximum theoretical transmission of around 30\% and is ideal for our SERS experiments,  but the intrinsically modest cavity Q is clearly not optimal for studies mapping cavity perturbation theory.

\begin{figure}
    \centering
    \includegraphics[width=0.85\textwidth]{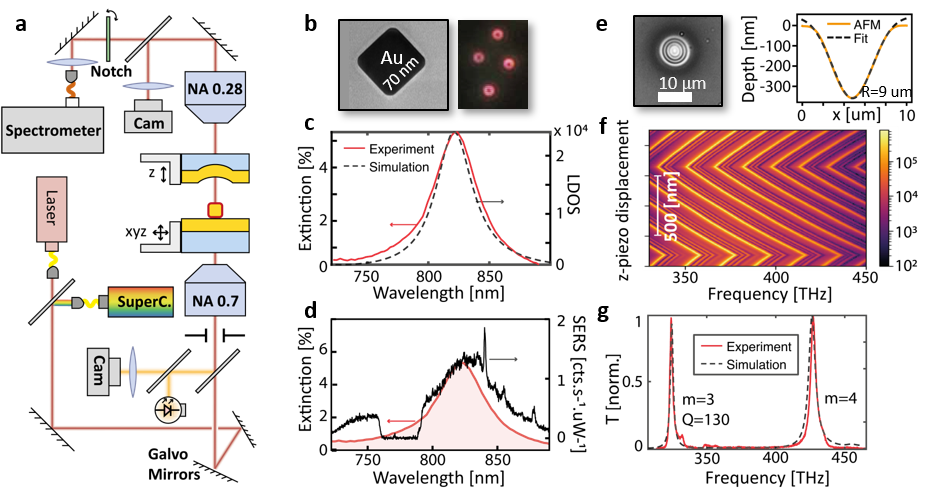}
    \caption{ (\textbf{a}) Home-built confocal Raman and white light spectroscopy setup working both in reflection and transmission. The sample position is controlled with two sets of piezos allowing cavity and NCoM alignment. (\textbf{b}) The NCoM is formed with 70 nm nanocubes deposited on a mirror surface and can be observed using dark-field imaging. (\textbf{c}) Extinction spectrum of a bare NCoM whose resonance frequency is tuned to 820 nm using a 6 nm alumina gap. (\textbf{d}) SERS spectrum taken on the same NCoM displays a broad background from electronic Raman of the gold and distinct peaks of BPT molecules that were previously coated on the nanocube. (\textbf{e-g}) A Fabry-Pérot cavity is formed with a second curved mirror fabricated using $\mathrm{CO}_2$ laser ablation which results in curved and smooth $\sim300$ nm deep concave features with a radius of curvature of about 9 \textmu m . The z displacement of the piezo allows tuning the cavity modes resonance (\textbf{f}) which feature Q in the range 100 to 300 depending on the mode order m as seen on a single transmission spectrum (\textbf{g}) .
    }
    \label{fig:Fab}
\end{figure}

Next we turn to experimental realization of the proposed platform. First, we fabricate flat mirrors by physical vapor deposition of 30~nm of Au on glass, over which we apply a thin spacer of Al$_2$O$_3$ (15 monolayers) using atomic layer deposition (ALD). To facilitate cavity alignment the  flat mirrors in our experiment are mesas carved out of a glass substrate. The NCoM system is formed by dropcasting 70~nm gold nanocubes (Fig.~\ref{fig:Fab}b) which have been previously functionalized with BPT molecules by self-assembly over a 12 hour incubation period.  For the second mirror forming the FP cavity, we use concave mirrors, fabricated by $\mathrm{CO}_2$ laser abblation of glass~\cite{Hunger2010,Barbour2011} and  subsequent coating by physical vapor deposition of 30 nm of gold. From atomic force microscopy (AFM) measurements of the fabricated concave features  we estimate an $\text{NA}=0.24$, a radius of curvature of $R=9$~\textmu m, and a roughness before coating of below 0.5~nm.

To interrogate the system we use a microscope setup that can perform measurements in transmission and reflection (Figure~\ref{fig:Fab}a).
At the heart of the setup is an assembly of micromechanical stages for positioning the curved mirror and the flat mirror relative to each other, and in the focus of the microscope objectives (XYZ and tip/tilt degrees of freedom). For the flat reflector, XYZ piezo actuation assures fine positioning of the cube, also allowing us to compare NCoM systems and flat mirror responses without modifying the set up. The curved reflector can be scanned such that we tune the cavity opening. Mechanical drifts of around 10 pm per second result in cavity frequency instabilities of $20/m$~pm/s, with $m$ the cavity mode order.
The setup  is equipped with a supercontinuum white light source  and with a narrowband tunable diode laser (Toptica DL-Pro 780, 50~nm tuning range) as Raman pump laser.  A galvanic mirror system allows for confocal imaging and fine positioning of the excitation spot.
We use a camera for positioning and alignment, and  a fiber-coupled Andor spectroscopy system (Andor Shamrock
A-SR-303i-B-SIL) equipped with a cooled silicon CCD camera (Andor iVac A-DR324B-FI) for spectroscopy. The Raman pump light is rejected with a set of two notch filters (Thorlabs NF785-33).
This setup allows for dark field scattering, cavity transmission, and SERS measurements --- all in both reflection and transmission.
Moreover, in the same setup we can compare cavity-NCoM performance with the bare NCoM system performance measured on the same nanocube. Finally, markers in the sample allow comparing individual cube data to scanning electron micrographs, and to extinction spectra that we take in a separate setup reported in \cite{Oksenberg2021}.

First, we plot the extinction spectrum of a bare NCoM in Figure~\ref{fig:Fab}c, obtained using a low NA objective. It shows the response to the supercontinuum laser, filtered with an AOTF (Crystal Technologies) over approximately 5 nm, whose center frequency is scanned. We obtain a resonance with $\text{Q}=17$ similar to what is expected from the simulated LDOS for the 11 mode. 
For the same cube, Figure~\ref{fig:Fab}d shows a bare NCoM SERS spectrum. The SERS spectrum (with the region around the pump wavelength blocked by a notch filter) shows a broad and featureless lineshape that is very close to the cube extinction resonance, and which we interpret as the SERS-enhanced electronic Raman signature of gold~\cite{Mertens2017}. Vibrational Raman lines of BPT appear on top of this background. The anti-Stokes signature is in this case less significant, due to the red detuning of the cube resonance relative to the pump wavelength. 
Figure~\ref{fig:Fab}e-g highlights properties of the cavity system in absence of any cube. In panel f we report transmission versus wavelength as we open and close the microcavity, using a saw-tooth voltage applied to the $z$-piezo.  A tapestry of cavity transmission resonances is clearly evident, with the main bright lines corresponding to fundamental modes of different longitudinal mode number. The fainter features in between the bright resonances are due to higher order transverse modes. This interpretation is confirmed by real-space imaging of the cavity output in transmission on a camera. Figure~\ref{fig:Fab}g reports transmission at a fixed cavity opening, showing that we can routinely access longitudinal  mode orders down to $m=3$ and above (up  to $m=10$ typically), with Q-factors scaling from 100 to 300.

\begin{figure*}
    \centering
    \includegraphics[width=0.9\textwidth]{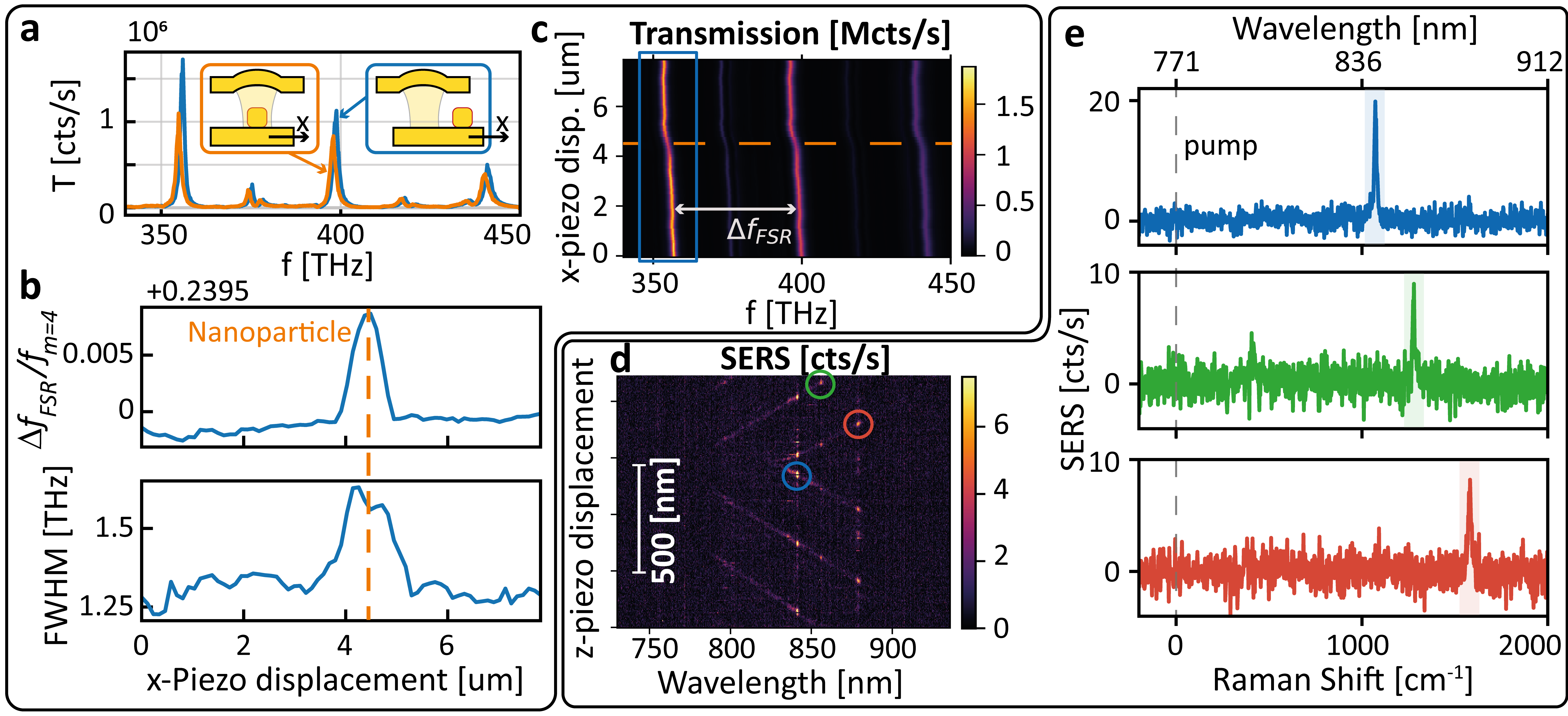}
    \caption{ (\textbf{c}) Cavity transmission when laterally displacing a nanoparticle from outside (empty cavity, blue curve in \textbf{a}) into the cavity space (hybrid resonator, orange curve in \textbf{a}). The nanoparticle induces a dispersive and dissipative shift (\textbf{b}) of the cavity resonance. The cavity frequency can be tuned by changing the distance between the mirrors. This allows selectively enhancing single Raman peaks of the BPT molecules (\textbf{d}). (\textbf{e}) Crosscuts corresponding to three main BPT Raman peaks.
    }
    \label{fig:sers1}
\end{figure*}

We can now assemble hybrid resonators and perform mode scatterometry and SERS. Figure~\ref{fig:sers1}a shows  transmission versus wavelength of a cavity of fixed geometry, with and without a metal nano-antenna. We accomplish this by sideways movement of the flat mirror, to scan the nano-antenna into and out of focus of the cavity.
We observe a small redshift of the cavity resonances, and evidence for a reduction in cavity Q, through a broadening and a reduction of the transmission peak height. As per Figure 2,  we indeed expect mainly a Q-factor reduction in the present operating regime. The Q drops approximately from $280$  to $220$, as expected for a $m=5$ mode and a hybrid coupling  strength of $J/2\pi=2\,\mathrm{THz}$.
The presence of the nanoparticle also results in a dispersive shift of the cavity frequency due to a modification of the optical path length. This is best visualized when plotting the shift of the FSR $\Delta f_\mathrm{FSR}$ normalized by the resonance frequency $f_m$ as shown in Figure~\ref{fig:sers1}b. Indeed $\Delta f_\mathrm{FSR}/f_m=1/m$ is a constant that does not depend on the cavity opening, and allows to get rid of the instabilities arising from the lateral piezo movement. Since for each cavity mode the frequency shift depends on the detuning with the nanoparticle, the normalized FSR is shifted when the cube is scanned laterally through the cavity. Both the linewidth increase and the dispersive shift are observed  over a lateral displacement distance of around 1 \textmu m. This distance is given by the lateral extent of the cavity mode at the mirror.

Figure~\ref{fig:sers1}d shows Raman spectra of the  sample where the pump laser wavelength is fixed at 771~nm,  and we modulate the cavity length back and forth. Generally, the vibrational lines of BPT, which should appear as vertical features in the diagram (since the laser is fixed), are hardly visible due to the fact that when used off-resonance, the cavity transmission function blocks the pump and the signal, making efficient pumping and signal collection impossible. Nonetheless, the fact that we scan the cavity is evident as faint diagonal stripes, indicative of enhancement of the electronic and molecular SERS by the fundamental and higher order cavity resonances.
Whenever a hybrid cavity resonance overlaps with a molecular line (e.g. at 841 nm for the blue circle), a bright burst of SERS signal is collected, indicating hybrid enhancement of the SERS signal.
Single crosscuts for SERS enhancement of three different individual vibrational lines are shown in panel e, and correspond to the spectra at the position of the bright spots marked by circles in panel d. They correspond to the 1079~$\mathrm{cm}^{-1}$, 1281~$\mathrm{cm}^{-1}$ and 1586~$\mathrm{cm}^{-1}$ BPT Raman lines respectively, and illustrate how the cavity allows for selective enhancement of the Raman signal by modifying the LDOS frequency landscape felt by the molecules.

\begin{figure*}
    \centering
    \includegraphics[width=\textwidth]{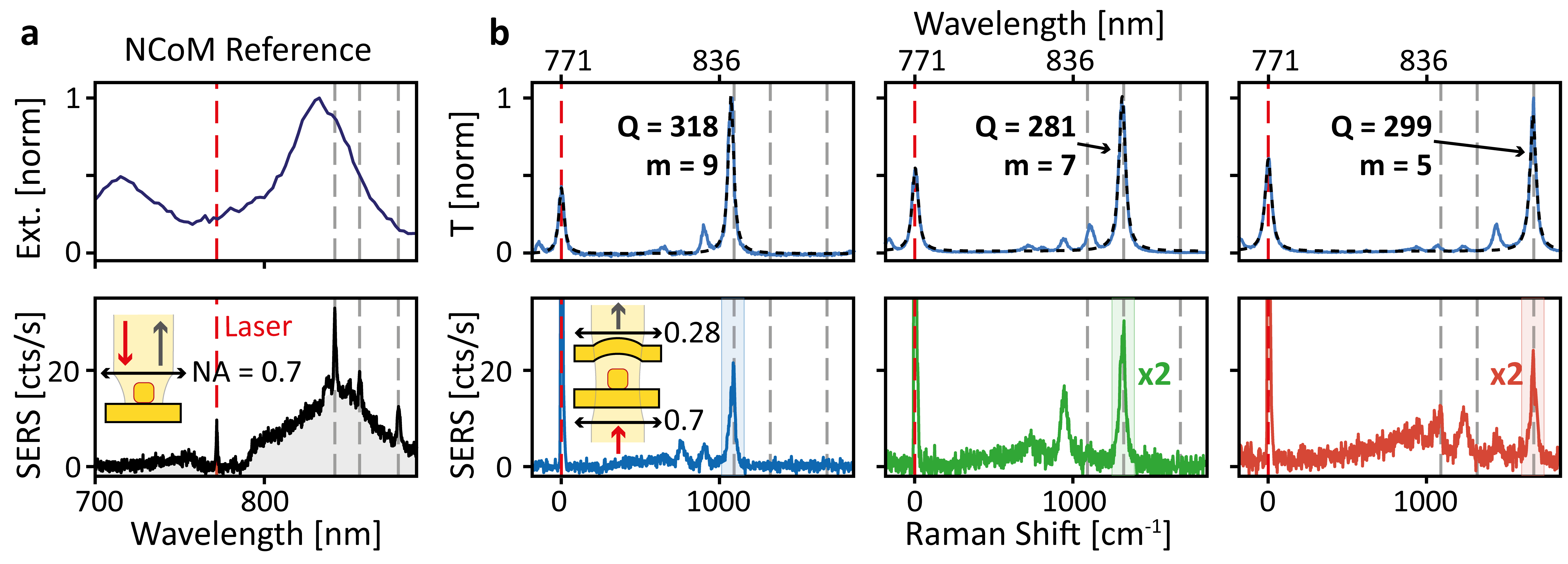}
    \caption{SERS spectra with (\textbf{a}) NCoM reference and (\textbf{b}) a hybrid resonator. The first row depicts WL characterization (extinction for the NCoM and transmission for the hybrid resonator), whereas the second row shows SERS spectra, where strong enhancement of the Raman signal is only obtained for molecular vibrational lines that are specifically tuned to the cavity resonance.}
    \label{fig:sers2}
\end{figure*}

Next our ambition is to optimize the use of the selective cavity-enhanced plasmonic SERS, and  to quantify the performance relative to standard NCoM SERS.  First, we realize that in Figure \ref{fig:sers1} the performance is limited because the system offers a cavity resonance at the Raman-shifted frequency,  but does not also offer a hybrid resonance for pump enhancement. Second, to quantify performance we perform quantitative count rate comparisons for cubes with and without  the closing cavity mirror. Figure \ref{fig:sers2} summarizes the results, with a direct correlative comparison white-light response of the resonator and the resulting SERS spectra.
The results for the reference NCoM (cube geometry verified on the SEM) without the top mirror are shown in the first column. The top panel shows the extinction spectrum featuring the broad Lorentzian resonance of the NCoM (11) mode. The SERS response shown in the bottom panel, taken with the $\text{NA}=0.95$ objective from the air side of the NCoM, follows closely this resonance and features both enhancement of gold electronic Raman and of the BPT lines. 
The remaining panels show cavity transmission on the top row, and the Raman spectra measured in transmission are shown in the bottom. For each column of graphs, one cavity mode is tuned to enhance the pump, while at the same time the FSR is tuned so that a second mode enhances the 1079, 1281 and 1586~$\mathrm{cm}^{-1}$ lines, successively. 
Not only  is sideband resolution obtained, with sideband resolution $\Omega_m/\kappa_\mathrm{eff}=25$ (where $\kappa_\mathrm{eff}$ is the cavity linewidth broadened by hybridization), but the relatively good Q of the cavity allows for specific vibrational line enhancement at will. All measurements are obtained for the same sample by simply varying the cavity opening. 
The Raman signal also shows enhancement of electronic Raman background from higher transverse order cavity modes that are only weakly visible in cavity transmission. They arise from slightly non-ideal behavior of the cavity (slightly tilted mirrors, non-Gaussian curvature) and from NCoM placement slightly away from the exact center of the cavity.

For both the hybrid resonator and the bare NCoM system, the same input laser power has been used; it is remarkable that essentially the same SERS count rates at given input power are obtained for the NCoM and the hybrid resonator, despite the fact that pumping and collection is now through two partially opaque metallic films and at a much lower NA (collection NA for the cavity transmission is 0.28). This eliminates the hypothesis of a cavity simply acting as a transmission filter for the Raman signal produced by the NCoM, since in that case, one would obtain at least a 20 times reduction in signal (5\% of cavity transmission measured at resonance) with an additional reduction related to mismatch between collection NA and NCoM emission pattern. 
Instead, the hybrid resonator allows us to reach  SERS enhancements in transmission and with a low NA that are equivalent to SERS enhancements obtained with bare NCoMs in optimal experimental settings, but with the important feature of working well beyond the sideband-resolved regime.

\section{Discussion}
We presented a novel tunable hybrid resonator,  consisting of a nanocube-on-mirror plasmonic antenna placed inside a Fabry-P\'erot cavity and benefiting from both the strong electromagnetic enhancement in the gap and the high-Q tunable resonance of the cavity. The resonance frequency of the hybrid mode can be tuned by changing the cavity opening, allowing enhancing specific Raman lines and placing the system well inside the sideband resolved regime. The hybridization has been first rationalized using a rigourous modal analysis and finite element modelling, which also showed the potential of this structure to study dissipative optomechanical coupling~\cite{Primo2020}. Experimental demonstration of in situ tunable sideband resolved SERS using gap plasmonic modes has been shown.
A quantitative comparison of the hybrid resonator and the bare nanoantenna resonator performance allowed by our bottom-up fabrication approach shows comparable SERS enhancements for both systems, and has been verified with a semi-analytical model.

Furthermore, further optimization is possible by considering the tuning of the  NCoM resonance wavelength.  We have chosen to work with NCoMs with resonances tuned between the laser pump and the Raman line wavelengths, not because it is optimal for the hybrid resonator, but because it is optimal for the bare nano-antenna system and also because it results in the strongest white-light transmission signatures of hybridization as shown in Fig.~\ref{fig:sers1}.  Thus this choice allows for a fair or even conservative comparison between the hybrid resonator and the nanoantenna.  Nonetheless,  even for large cavity-antenna detunings (on the order of a few antenna linewidths) the hybridized cavity mode can sustain high LDOS and field enhancement values on par with the maximum values obtained for the bare NCoM ~\cite{Doeleman2016}.
This unique property can be used to maintain sideband resolution with the cavity mode, whilst reducing undesired photon scattering into the broad antenna mode by working with a far detuned antenna with respect to the pump and the Raman lines. This can for example allow to strongly enhance the Stokes process while avoiding most of the anti-Stokes scattering, leading to phonon build-up and non-linear Raman scattering~\cite{Roelli2016}. For example, by red-detuning the antenna by two antenna linewidths ($\omega_L-\omega_a=2\gamma$) the total anti-Stokes scattering can be divided by a factor 40 compared to the resonant antenna case, while the hybrid resonator with the parameters obtained for silver mirrors allows to reach Stokes scattering with a rate beyond 50\% of the resonant antenna (see supplementary). Inversely, in the opposite configuration the hybrid can be used to strongly suppress the Stokes process which is readily useful to reduce the laser-related noise in mid-IR upconversion schemes~\cite{Roelli2020,Chen2021,Xomalis2021}.

Overall, the optomechanical cooperativity of the hybrid resonator is just limited by the performance of the antenna mode that is used, and by reducing the gap size of the nanocube-on-mirror system (which would advantageously red-detune the antenna mode at the same time) one could tend towards optomechanical cooperativities of order one which are necessary to observe non-linear effects~\cite{Benz2016Single-moleculepicocavities,Xu2022}.   
The performance of the system can readily be boosted by inserting multiple nanoparticles in the larger active area of the hybrid cavity mode, and presents an interesting perspective to study long-range multimodal molecular optomechanics with hybrids.


\appendix
\section{Materials and methods}
\subsection{Semi-analytical model for SERS enhancement}
\begin{figure*}
    \centering
    \includegraphics[width=0.5\textwidth]{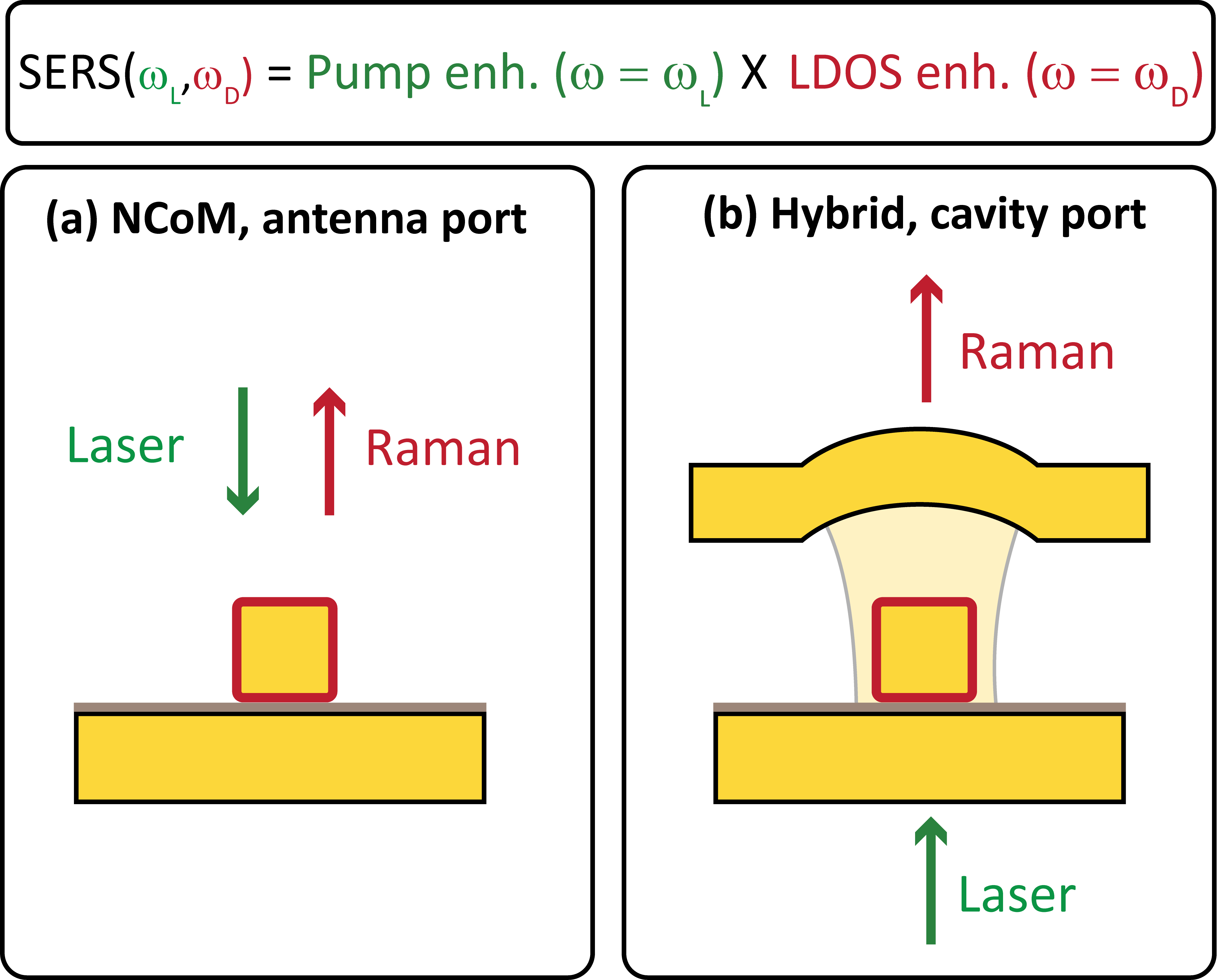}
    \caption{ Both processes involved in the parametric Raman scattering can be enhanced to reach high SERS enhancements. For multimode systems the enhancement depends on the specific optical channel that is addressed~\cite{Shlesinger2021}. The two different configurations that are compared in the main text: (\textbf{a}) the NCoM reference and (\textbf{b}) the NCoM in a Fabry-P\'erot hybrid. The former involves the antenna optical channel for excitation and emission, the hybrid resonator involves the cavity channel for both processes.
    }
    \label{fig:app0}
\end{figure*}

We recall here the main derivation steps of the model to calculate the SERS enhancement of a multimodal resonator with multiple optical input and output ports that has been presented elsewhere~\cite{Shlesinger2021}. It relies on the derivation of the semi-classical Langevin equations for each of the coupled optical and mechanical resonators.
Raman enhancement writes as a product of pump enhancement $\mathrm{PE}$ and collected local density of optical states $\mathrm{LDOSC}$. 
Both depend on the optical channel that is being used (which differs depending on the considered resonator), and on the wavelength of the corresponding process (laser wavelength for pump enhancement, and Raman line wavelength for the LDOS enhancement) as sketched in Figure~\ref{fig:app0}.
The pump enhancement is written as the ratio of the electric field intensity with and without the resonator. For the hybrid resonator it writes:
\begin{equation}\label{eq:pump_enh}
    \text{Pump enh.}=\left|\frac{\bar{\alpha}_a\tilde{E}_a+\bar{\alpha}_c\tilde{E}_c}{E_{\mathrm{inc}}}\right|^2,
\end{equation}
with $\tilde{E}_{a,c}(\mathbf{r_0})$ the normalized mode profile of the antenna or cavity (normal modes are considered in the model), evaluated at the position of the molecule, and related to the (real) mode volume by $V_{a,c}= \frac{2}{\epsilon_0 \epsilon |\tilde{E}_{a,c}(\mathbf{r_0})|^2}$, $E_\mathrm{inc}$ is the incident electric field at the focus spot, and $|E_\mathrm{inc}|^2$ the incident power.  
$\bar{\alpha}_{a,c}$ are the steady state amplitude of the antenna or cavity optical modes:
\begin{align}
\begin{cases}
\bar{\alpha}_a=\chi_a'\left(\sqrt{\eta_{\text{in},a}\gamma_{\mathrm{rad}}}\,S_{\text{in},a}+iJ^*\chi_c \sqrt{\eta_{\text{in},c}\kappa}\,S_{\text{in},c}\right)\\
\bar{\alpha}_c=\chi_c'\left(\sqrt{\eta_{\text{in},c}\kappa}\,S_{\text{in},c}+iJ\chi_a \sqrt{\eta_{\text{in},a}\gamma_{\mathrm{rad}}}\,S_{\text{in},a}\right)\,,
\end{cases}\label{eq:avgsol}
\end{align}
where $\eta_\mathrm{in,a,c}$ are the input efficiencies and $S_{\text{in},a,c}$ the input pump amplitude in each port, such that $|S_{\text{in},a,c}|^2$ is the incoming optical pump power.
The susceptibilities of the bare cavity mode $\chi_c$ and antenna mode $\chi_a$ are, in the rotating wave approximation,
\begin{align}
\begin{cases}
    \chi_a(\omega)=\frac{i}{\omega-\omega_a+i\frac{\gamma_a}{2}}\,,\\
    \chi_c(\omega)=\frac{i}{\omega-\omega_c+i\frac{\kappa}{2}}\,,
\end{cases}\,
\end{align}
and the hybridized cavity of hybridized antenna susceptibilities $\chi_{c,a}'$ read
\begin{equation}\label{eq:suscept-hybrid}
     \chi_{c,a}'(\omega)=\frac{\chi_{c,a}}{1+|J|^2\chi_a\chi_c}.
\end{equation}
The LDOSC is given as a function of the resonator susceptibility and writes for a collection through the cavity port: 
\begin{align}\label{eq:LDOSac}
    \mathrm{LDOSC}_\mathrm{cav}(\omega)&=\eta_{\mathrm{out,c}}\kappa\frac{3\pi\epsilon_0 c^3}{2n^3\omega^2} \times \nonumber\\
    &\lvert\chi_c'(\omega)\left(\tilde{E}_c^*+iJ^*\chi_a(\omega)\tilde{E}_a^*\right)\rvert^2\, ,
\end{align}
where $\eta_{\mathrm{out,c}}$ is the output coupling efficiency, $n$ is the refractive index at the molecule position and $\chi_c'$ and $\chi_a$ are the hybridized cavity susceptibility and antenna susceptibility.

For the NCoM (reference case), there is only the antenna mode so $\chi_c=0$ and $\chi_a'=\chi_a$. The only port for input and output is the antenna optical channel. So we can write the respective pump enhancement and collected LDOS as:
\begin{align}
    \mathrm{PE}_{\mathrm{NCoM}} &= \eta_{\mathrm{in},a}\gamma \frac{S_{\mathrm{inc},a}}{|E_{\mathrm{inc},a}|^2}  |\chi_a|^2\frac{2}{\epsilon_0 \epsilon V_a}\\ \nonumber
    \mathrm{LDOSC}_{\mathrm{NCoM}} &= \eta_{\mathrm{out},a}\gamma\frac{3\pi c^3}{2 n^3\omega^2}\frac{|\chi_a|^2}{V_a}
\end{align}
The susceptibility of the antenna needs to be evaluated at the pump frequency for the pump-enhancement case and at the emission frequency for the LDOS.

In our case, and what is ideal in a usual Raman situation, the antenna is approximately in the middle of the pump and emission frequencies, i.e. $\omega_a\simeq1/2(\omega_L+\omega_D)$, so the susceptibility of the antenna modulus will be the same in both cases. Hence, the frequency is omitted in the following.
For the hybrid resonator the input and output is through the cavity port:
\begin{align}
    \label{eq:hybLDOS}
    \mathrm{PE}_{\mathrm{hyb}} &= \eta_{in,c}\kappa  \frac{S_\mathrm{inc, c}}{|E_{\mathrm{inc,c}}|^2}  |\chi_c'|^2\frac{2}{\epsilon_0 \epsilon V_c}\left|1+ iJ\chi_a\sqrt{\frac{V_c}{V_a}}\right|^2 
    \simeq \eta_{in,c}\kappa  \frac{S_\mathrm{inc, c}}{|E_{\mathrm{inc,c}}|^2}  |\chi_c'|^2\frac{2}{\epsilon_0 \epsilon V_a}\left|J\chi_a\right|^2 
    \\ \nonumber
    \mathrm{LDOSC}_{\mathrm{hyb}} &= \eta_{out,c}\kappa \frac{3\pi c^3}{2 n^3 \omega^2}\frac{|\chi_c'|^2}{V_c} \left| 1+ iJ^*\chi_a\sqrt{\frac{V_c}{V_a}}\right|^2 
    \simeq \eta_{out,c}\kappa \frac{3\pi c^3}{2 n^3\omega^2}\frac{|\chi_c'|^2}{V_a} \left|J^*\chi_a\right|^2 
    \,.
\end{align}
For the hybrid, one cavity mode is resonant with the laser frequency for the pump enhancement and another is resonant with the Raman line for the LDOS enhancement, so the cavity susceptibility can be simplified each time as $\chi_c(\omega_c)=2/\kappa$ and the hybridized susceptibility as
\begin{equation}\label{eq:suscept-hybrid}
     \chi_c'(\omega=\omega_c)=\frac{2}{\kappa+2|J|^2\chi_a(\omega_c)} = \frac{2}{\kappa}\frac{1}{1+C_\mathrm{hyb}(1-i\tilde{\Delta}_a)^{-1}},
\end{equation}
with 
\begin{equation}
    C_\mathrm{hyb} = \frac{4J^2}{\kappa\gamma}\,
\end{equation}
the hybrid cooperativity and $\tilde{\Delta}_a=2(\omega-\omega_a)/\gamma$.

As such, comparing the two enhancements we obtain
\begin{align}\label{eq:SERSenh}
    \frac{\mathrm{SERS}_{\mathrm{hyb}}}{\mathrm{SERS}_\mathrm{NCoM}} &= \frac{\eta_{in,c}}{\eta_{in,a}} \frac{\eta_{out,c}}{\eta_{out,a}} \left(\frac{\mathrm{NA}_a}{\mathrm{NA}_c} \right)^2\left(\frac{\kappa}{\gamma} \right)^2\left|\frac{\chi_c'}{\chi_a} \right|^4 \left|\frac{V_a}{V_c}\right|^2\left| 1+iJ\chi_a\cdot\sqrt{\frac{V_c}{V_a}}\right|^4 \\
    \frac{\mathrm{SERS}_{\mathrm{hyb}}}{\mathrm{SERS}_\mathrm{NCoM}} 
    &\simeq \frac{\eta_{in,c}}{\eta_{in,a}} \frac{\eta_{out,c}}{\eta_{out,a}} \left(\frac{\mathrm{NA}_a}{\mathrm{NA}_c} \right)^2  \frac{C_\mathrm{hyb}^2}{\left|1+C_\mathrm{hyb}(1-i\tilde{\Delta}_a)^{-1}\right|^4 }
    \,,
\end{align}
where the last equation is obtained by using the approximated hybrid SERS expression of Eq.~\ref{eq:hybLDOS}.
We have also used $|\chi_a(\omega_L)|=|\chi_a(\omega_D)|$ (antenna optimally placed in between pump and Raman frequency), and the fact that we use two cavity modes; one tuned to the pump and the other to the emission. For a large antenna detuning (i.e. $\gamma\ll\omega_a-\omega_{D,L}$ and hence $\tilde{\Delta}_a\gg1$), the hybrid resonator can strongly surpass the Raman enhancement of the antenna. This is due to the fact that the hybrid resonator can enhance simultaneously the pump and the Raman line (with two distinct cavity modes) whereas the single mode antenna cannot. 
For the values gathered in table~\ref{table:1} describing the experimental conditions, we expect a ratio of enhancements of order 1.
\begin{table}[h!]
\centering
\begin{tabular}{||c c c||} 
 \hline
 Parameter & Value & Comment \\ [0.5ex] 
 \hline\hline
 $\lambda_P$ & 771 nm & laser pump \\ 
 $\lambda_D$  & 840 nm & 1st BPT Raman line \\
$\lambda_{NCoM}$ & 821 nm & from extinction \\
$\lambda_c$ &  $\lambda_P$ or  $\lambda_D$ &  cavity resonant with pump and BPT peak \\
 $n$ & 1.76 & alumina refractive index \\
 $\eta_{in,a}$ & 1.25\% & from extinction and $\gamma_\mathrm{rad}/\gamma$ \\
 $\eta_{out,a}$ & 12\% & $\mathrm{asin}(\mathrm{NA}_a)/(\pi/2) \cdot \gamma_\mathrm{rad}/\gamma$ \\
 $\eta_{out,c}$ & 10\% &  $T/(2\cdot(1-R))$ with T and R transm. and reflec. of Au mirror \\
 $\eta_{in,c}$ & 18\%  & given by mode matching of cavity and input objective NA \\
$Q_c$, $Q_a$ & 22 , 280 & from comsol / experiment (m=10) \\ 
$\gamma_\mathrm{rad}/\gamma$ & 1/4 & 1/2 in Joules, 1/4 in plasmon emission \\
$V_a$ & $1.3\cdot10^{-5}\lambda^3$ & vertical electric dipole 1 nm below corner \\
$V_c$ & $34.4\cdot\lambda^3$ & vertical electric dipole 1nm below corner (poorly matched to cavity mode)\\
$J/2\pi$ & 3.4 THz & for m=2, from fit of LDOS 2D map with hybrid model \\ [1ex] 
 \hline
\end{tabular}
\caption{Parameters used in the semi-analytical model.}
\label{table:1}
\end{table}
In Figure 5 we present ratios of detected counts instead of SERS enhancements. Equation \ref{eq:SERSenh} is transposed to counts by multiplying the enhancement rate by the spectra obtained for a Raman dipole in vacuum $S_\mathrm{ref,a,c}\propto|E_\mathrm{inc,a,c}|^2$. For the same input power (laser power before the input objective), the electric field at the sample position $E_\mathrm{inc,a,c}\propto1/\mathrm{NA_{a,c}}$ so the only change to Eq.\ref{eq:SERSenh} will be a drop of the NA factors.

\subsection{Determining the hybrid coupling parameter $J$}
\begin{figure*}
    \centering
    \includegraphics[width=0.9\textwidth]{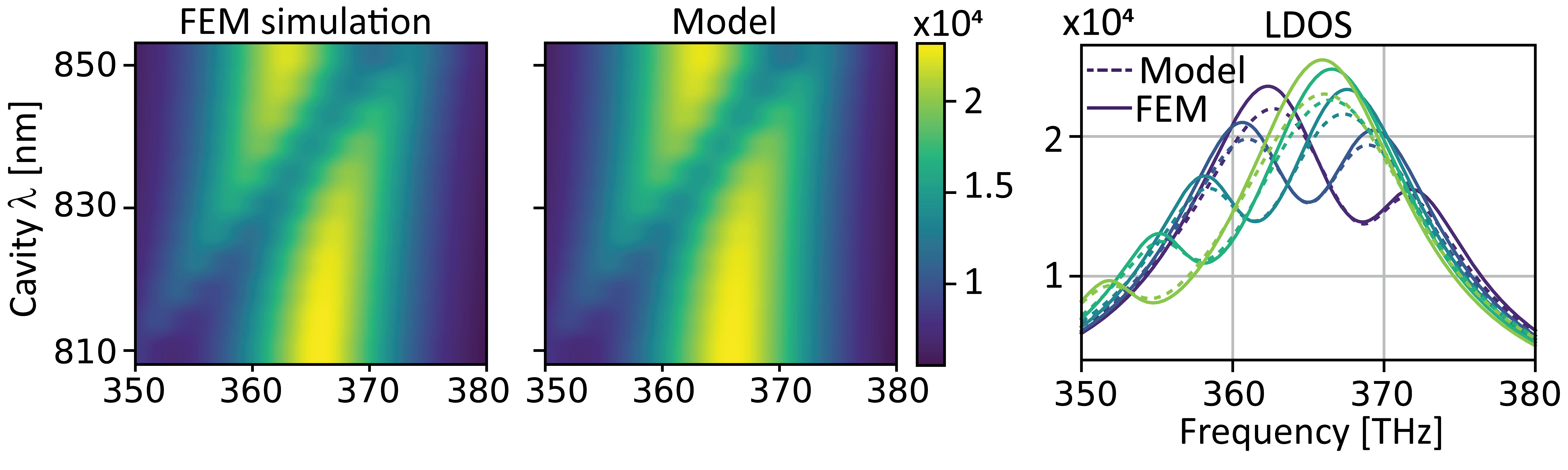}
    \caption{ Fitting the LDOS obtained with FEM simulations with a semi-analytical model. Only J is left as a free parameter, and the complete 2D map is fitted simultaneously. We obtain $J/2\pi=3.4\,\mathrm{THz}$. Right plot: crosscuts of the 2D map for different cavity openings, showing a residual smaller than 10\%.
    }
    \label{fig:app1}
\end{figure*}

\begin{figure*}
    \centering
    \includegraphics[width=0.9\textwidth]{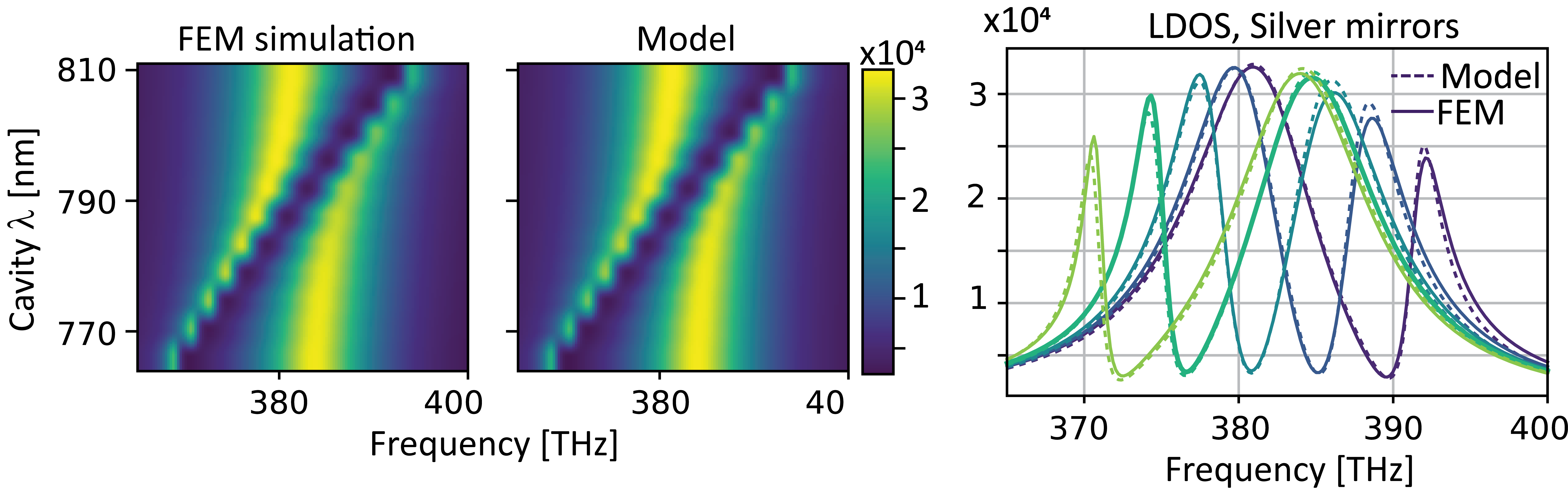}
    \caption{ Same as figure~\ref{fig:app1} but using 50 nm silver mirrors for the cavity instead. The bare cavity Q factor for the m=2 is then around Q=600, which results in better Fano contrast and higher LDOS for the cavity-like mode.
    }
    \label{fig:app2}
\end{figure*}

The electromagnetic interaction between the two optical modes is modelled by a hybrid coupling rate $J$. In the case of a dipolar antenna coupled to a single cavity mode it can be expressed as a function of the bare antenna polarizability and the bare cavity mode volume~\cite{Doeleman2016}. Here the interaction involves a magnetic dipole gap mode, and $J$ can be retrieved using the magnetic polarizability and mode volume. For simplicity, we choose to fit the analytical model using the simulated LDOS map obtained with COMSOL and the rigourous modal analysis toolbox MAN~\cite{Yan2018}. The only free parameter is the hybrid coupling rate J, all other parameters are extracted from experimental values or simulated results obtained for the bare resonators. 
The complete 2D map (as a function of cavity opening and detected frequency) is fitted simultaneously. The result is shown in Figure~\ref{fig:app1}.
Since the FEM calculation gives the total LDOS we compute the total LDOS from the model as a sum of hybridized cavity mode LDOS (Eq.~\ref{eq:hybLDOS}) and hybridized antenna LDOS:
\begin{align}
    \mathrm{LDOS}_{\mathrm{hyb,c}} &= \kappa \frac{3\pi c^3}{2 n^3\omega^2}\frac{|\chi_c'|^2}{V_c} \left| 1+ iJ^*\chi_a\sqrt{\frac{V_c}{V_a}}\right|^2  \\    
    \mathrm{LDOS}_{\mathrm{hyb,a}} &= \gamma \frac{3\pi c^3}{2 n^3\omega^2}\frac{|\chi_a'|^2}{V_a} \left| 1+ iJ^*\chi_c\sqrt{\frac{V_a}{V_c}}\right|^2 
    \,.
\end{align}
The performance of the hybrid resonator is mainly limited by the poor quality factor of the 30~nm thick gold mirrors chosen to facilitate the experimental proof of concept. By simply using 50~nm silver mirrors instead, we obtain a cavity Q that is one order of magnitude higher, and an improved LDOS for the cavity-like hybrid mode as shown in Fig.~\ref{fig:app2}.
    
\subsection{Sideband resolution in multimode molecular optomechanics}
The use of narrow optical modes allows tailoring the phase of the optomechanical interaction, which allows enhancing either the Stokes or the anti-Stokes process alone. In single-mode optomechanics this is quantified by the sideband resolution ratio $\Omega_m/\kappa$, where $\kappa$ is the linewidth of the optical resonance. Plasmonic resonances with a typical quality factor $Q\sim10$ cannot efficiently enhance one sideband without resulting on a significant enhancement of the opposite sideband due to the tail of the broad plasmonic mode. 
By using multimode hybrid resonators, the plasmonic resonance can be strongly detuned to completely suppress the unwanted scattering process, while still enhancing the desired sideband by using the narrow hybridized cavity mode. The sideband resolution in this case can be quantified by comparing the total scattering rate of each sideband. For example, parametric amplification necessitates a strong enhancement of the Stokes scattering rate $A^+$ and a suppression of anti-Stokes phonon anihilation rate $A^-$\cite{Aspelmeyer2014CavityOptomechanics}. For a hybrid resonator the ratio of the two writes:
\begin{equation}
    \frac{A^+}{A^-} = \left( \frac{C_\mathrm{hyb}}{\frac{4\Delta_a^2}{\gamma^2}+C_\mathrm{hyb}} \right)^2\cdot\left(1+\frac{2(\Delta_a+\Omega_m)}{\gamma}\right)^2\, .
\end{equation}
For a well detuned antenna, and good hybrid cooperativities, this ratio diverges. For example with 50 nm silver mirrors resulting in $C_\mathrm{hyb}=42$ and an antenna detuned by twice the linewidth, we get a ratio of $A^+/A^-\simeq20$. This is considering a mechanical frequnecy $\Omega_m = \gamma$. The hybrid can still maintain sideband resolution even for very small Raman shifts $\Omega_m\ll\gamma$, with a ratio that tends to 10 in this case.

\subsection{Cavity parameters} 
\subsubsection{Cavity mode volumes and Q vs m}

The NA corresponding to the gaussain mode of a FP cavity is given by:
\begin{equation}
    \mathrm{NA} = \frac{\lambda}{\pi w_0}\,, 
\end{equation}
where the waist $w_0$ is related to the radius of curvature $R$ of the mirror and the length of the cavity $L$ by:
\begin{equation}
    w_0 = \sqrt{\sqrt{\frac{R}{L}-1}\frac{\lambda L}{\pi} }\,.
\end{equation}
For $R=9$~\textmu m and typical $L=2$~\textmu m we get an NA for the cavity mode of 0.26.

\subsubsection{Fabrication}
For the NCoM:
A quartz substrate is pre-treated, where a platform elevated from the rest of the substrate by a few hundred microns is created using a water jet and facilitates the cavity alignment.
After that, electron-beam evaporation is performed using a Polyteknik E-Flex, depositing 30~nm Au at a deposition rate of 0.5~\AA/s. A 2~nm adhesion layer of Cr is used.
The thickness of the gold layer is a compromise between visibility of the cubes in microscopy, and the quality of the plasmonic cavity. 
An Al$_2$O$_3$ spacer layer of variable thickness is grown on top. We dropcast nanocubes (75~nm nominal, Nanopartz), incubated for 12 hours in a 1~mM biphenyl-4-thiol (BPT) solution in anhydrous ethanol to form a self-assembled molecular monolayer,  and perform darkfield microscopy to verify successful assembly of the NCOM system. 
The typical single-cube radiation pattern in darkfield microscopy, observed for a spacer of 6 nm is shown in Fig.~\ref{fig:Fab}(c). The donuts are observed for a high-NA dark field objective (collecting only signal from large angles, i.e. the 10 mode). For lower NA, we collect the magnetic mode 11, which gives filled round shapes.

For the curved mirror: we use CO2 laser ( Synrad 48-1 working at $\lambda=10.6\,\mathrm{\mu m}$) ablation to create concave mirrors.  We use a home-built set up in which we focus a pulse of 100 ms and 300 mW to a tight spot using an IR lens (black diamond, NA=0.5, Thorlabs). We tune the exposure dose and focusing to obtain mirrors with a radius of curvature of around 9 microns, and a depth of 350 nm. These mirrors can subtend an NA of 0.25.  We verified the mirror topography by AFM, observing a roughness below 0.5 nm.

\begin{acknowledgments}  
The authors thank Isabelle Palstra, Beniamino Ferrando, Kevin Cogn\'ee and Said Rodriguez for fruitful discussions.
This work is part of the Research Program of the Netherlands Organization for Scientific Research (NWO). The authors acknowledge support from the European Unions Horizon 2020 research and innovation program under Grant Agreements No. 829067 (FET Open THOR) and No. 732894 (FET Proactive HOT). 
\end{acknowledgments}

%

\end{document}